\documentclass[a4paper,11pt]{article}
\usepackage{pos}
\usepackage{placeins}
\usepackage{multirow}
\usepackage{braket}

\usepackage[font=small, labelfont=bf, textfont={small,it}]{caption}
\title{The Twisted Gradient Flow strong coupling with Parallel Tempering on Boundary Conditions}
\ShortTitle{The TGF strong coupling with PTBC}

\author[a]{Claudio Bonanno}
\author*[a,b]{Jorge Luis Dasilva Gol\'an}
\author[c]{Massimo D'Elia}
\author[a]{Margarita Garc\'ia P\'erez}
\author[c]{Andrea Giorgieri}

\affiliation[a]{Instituto de F\'isica T\'eorica UAM-CSIC, Calle Nicol\'as Cabrera 13-15,\\Universidad Aut\'onoma de Madrid, Cantoblanco, E-28049 Madrid, Spain}

\affiliation[b]{Departamento de F\'isica Te\'orica, Universidad Aut\'onoma de Madrid,\\M\'odulo 15, Cantoblanco, E-28049 Madrid, Spain}

\affiliation[c]{Dipartimento di Fisica dell'Universit\`a di Pisa \& \\ INFN Sezione di Pisa, Largo Pontecorvo 3, I-56127 Pisa, Italy\\\vspace{\baselineskip}}

\emailAdd{claudio.bonanno@csic.es}
\emailAdd{jorge.dasilva@uam.es}
\emailAdd{massimo.delia@unipi.it}
\emailAdd{margarita.garcia@uam.es}
\emailAdd{andrea.giorgieri@phd.unipi.it}

\abstract{We present a proposal for calculating the running of the coupling constant of the $\mathrm{SU}(3)$ pure-gauge theory, which combines the Twisted Gradient Flow (TGF) renormalization scheme with Parallel Tempering on Boundary Conditions (PTBC). The TGF is a gradient flow-based renormalization scheme formulated in an asymmetric lattice with twisted boundary conditions. Combined with step scaling, it has been successfully used to calculate the $\mathrm{SU}(3)$ $\Lambda$ parameter. As with all gradient flow-based schemes, the coupling constant is highly correlated with the topological charge and affected by topology freezing, an issue addressed by projecting the determination of the coupling onto the zero topological sector. As an alternative to the zero-charge projection, we combine TGF with PTBC, by replicating multiple copies of the same lattice, interpolating between periodic and open boundary conditions in a parallel-tempered manner. We present a first exploration of these ideas by analyzing specific ensembles of $\mathrm{SU}(3)$ lattices with and without PTBC.}

\FullConference{The 40$^{\text{th}}$ International Symposium on Lattice Field Theory (Lattice 2023)\\
July 31$^{\text{st}}$ - August 4$^{\text{th}}$, 2023\\
Fermi National Accelerator Laboratory\\}

\newcommand{\beq}{\begin{eqnarray}}
\newcommand{\eeq}{\end{eqnarray}}
\newcommand{\beqnn}{\begin{eqnarray*}}
\newcommand{\eeqnn}{\end{eqnarray*}}

\newcommand{\SU}{\mathrm{SU}}
\newcommand{\ov}{\mathrm{ov}}
\newcommand{\TGF}{\mathrm{TGF}}

\newcommand{\strong}{\mathrm{strong}}
\newcommand{\MSbar}{\overline{\mathrm{MS}}}
\newcommand{\Tr}{\mathrm{Tr}}
\newcommand{\f}{\mathrm{f}}
\newcommand{\tl}{\tilde l}
\newcommand{\tL}{\tilde L}

\begin{document}
\maketitle

\section{Introduction}

Recently, there was a renewed interest in the lattice numerical calculation of the renormalized running coupling $\lambda$ of the pure-gauge $\SU(3)$ Yang--Mills theory (i.e., with $N_{\f}=0$ quark flavors) and of its related dynamically-generated scale $\Lambda_{\MSbar}^{(0)}$, as this quantity is necessary to obtain a determination of the coupling $\alpha^{(N_\f)}_{\strong}$ of the full theory with $N_\f$ dynamical fermions via the decoupling method~\cite{DallaBrida:2019mqg}, see also Ref.~\cite{DelDebbio:2021ryq} for a recent review. Lattice determinations of $\Lambda_{\MSbar}^{(0)}$ have been constantly refined over the last 10 years~\cite{Brambilla:2010pp,Asakawa:2015vta,Kitazawa:2016dsl,Ishikawa:2017xam,Husung:2017qjz,DallaBrida:2019wur,Nada:2020jay,Husung:2020pxg,Bribian:2021cmg,FlavourLatticeAveragingGroupFLAG:2021npn} and going beyond the present state of the art requires a solid control over several possible sources of systematic errors.

An issue which could potentially introduce undesired systematic effects in the computation of the renormalized coupling is the relation of the latter quantity with topology. As a matter of fact, it is customary to compute the renormalized coupling from the action density via the gradient flow~\cite{Narayanan:2006rf,Lohmayer:2011si,Luscher:2009eq}; however, it is well known that, after the flow, such quantity is highly correlated with the background topological charge of the gauge fields~\cite{Fritzsch:2013yxa}. The topological charge in turn suffers for long auto-correlation times, rapidly increasing as one approaches the continuum limit and leading to the infamous \emph{topological freezing} problem~\cite{DelDebbio:2004xh}. 
This prevents a proper sampling of the topological charge distribution unless an unreasonable amount of statistics is collected. Given the strong correlation between the charge and the action density, it is thus clear that topological freezing can potentially introduce a sampling problem in the action density too; therefore, in order to avoid undesired systematics, it is important to check that topological freezing has no significant impact on the determination of the coupling from the gradient flow.

In most cases, this issue is bypassed by computing the coupling projecting the action density in the $Q=0$ topological charge sector, which is the only expected relevant topological sector in the perturbative regime; this amounts to a particular choice of renormalization scheme~\cite{Fritzsch:2013yxa}. However, there is in principle no guarantee that, in the presence of large auto-correlation times of $Q$, the space of gauge configurations is correctly sampled within each individual topological sector. Thus, the question whether topological freezing has an impact on the determination of the coupling via the gradient flow still remains valid even when considering the projected coupling.

In this proceedings, we present a preliminary investigation of the impact of topological freezing on the determination of the gradient flow coupling based on the adoption of a recently-proposed algorithm, the \emph{Parallel Tempering on Boundary Conditions} (PTBC). This algorithm, initially introduced and adopted for $2d$ $\mathrm{CP}^{N-1}$ models~\cite{Hasenbusch:2017unr,Berni:2019bch,Bonanno:2022hmz}, was in recent times successfully applied to $\SU(N)$ gauge theories too, achieving impressive reductions of the auto-correlation time of the topological charge also in this case~\cite{Bonanno:2020hht,Bonanno:2022yjr}. The renormalized coupling was instead computed within the so-called \emph{Twisted Gradient Flow} (TGF) scheme~\cite{Ramos:2014kla,Bribian:2019ybc,Bribian:2021cmg}, which is based on two main ingredients: the adoption of Twisted Boundary Conditions (TBCs)~\cite{tHooft:1979rtg,tHooft:1980kjq}, and the use of a peculiar geometry for the lattice, namely $L^2 \times \tL^2$, with $\tL=NL$ and $N$ the number of colors. The latter choices are motivated by the idea of volume reduction in the presence of TBCs~\cite{Gonzalez-Arroyo:1982hyq,Gonzalez-Arroyo:1982hwr,Gonzalez-Arroyo:2010omx} (see, e.g., Ref.~\cite{GarciaPerez:2014cmv,GarciaPerez:2020gnf} for a review and further references). As a matter of fact, it can be shown that with our setup the lattice has effectively the dynamics of a $\tL^4$-site lattice.

This manuscript is organized as follows: in Sec.~\ref{sec:alg} we briefly describe our PTBC algorithmic setup, in Sec.~\ref{sec:res} we present our preliminary results for the renormalized TGF coupling obtained with the PTBC algorithm, finally in Sec.~\ref{sec:conclu} we draw our conclusions.

\section{The PTBC algorithm}\label{sec:alg}

We discretize the pure-gauge Yang-Mills action on a $L^2 \times \tL^2$ lattice ($\tL=NL=3L$) using the standard Wilson action. By convention, we label the short directions with $\mu=1,2$. We consider $N_r$ replicas of this lattice, each one differing for the boundary conditions imposed on a small sub-region, referred to as the \emph{defect}, chosen in order to interpolate between periodic and open boundary conditions. Away from the defect we impose periodic boundary conditions everywhere except for some plaquettes lying in the short plane, which are instead multiplied by a phase factor, implementing TBCs. In practice, the lattice action of a given replica, labeled with the index $r$, reads:
\beq
S_{\mathrm{W}}^{(r)}[U] = -\frac{\beta}{N}\sum_{x,\mu>\nu} K_{\mu\nu}^{(r)}(x) Z_{\mu\nu}^*(x)\Re\Tr \left[ P_{\mu\nu}(x)\right],
\eeq
where $\beta=2N/g^2$ is the (inverse) bare coupling; $P_{\mu\nu}(x) = U_\mu(x) U_\nu(x+a\hat{\mu}) U_\mu^{\dag}(x+a\hat{\nu}) U^{\dag}_\nu(x)$ is the product of the links along the plaquette rooted in the site $x$ and lying in the $(\mu,\nu)$ plane; the numerical factor $Z_{\mu\nu}(x)$, used to impose TBCs, is equal to one for all plaquettes except for those with sites $x_1=x_2=1$ and lying in the short plane $(\mu,\nu)=(1,2)$, for which $Z_{12}=Z_{21}^*=\exp\{i 2 \pi / N\}$; finally, the numerical factor $K_{\mu\nu}^{(r)}(x) = K_{\mu}^{(r)}(x)K_{\nu}^{(r)}(x+a\hat{\mu})K_\mu^{(r)}(x+a\hat{\nu})K_\nu^{(r)}(x)$ is used to impose varying boundary conditions on the defect. As a matter of fact, the quantity $K_{\mu}^{(r)}(x)$, attached to each link, is equal to 1 unless the link $U_\mu(x)$ crosses orthogonally the defect, in which case $K_{\mu}^{(r)}(x)=c(r)$, where $0\le c(r) \le 1$. The extreme values $c(r)=1$ and $c(r)=0$ correspond respectively to periodic boundary conditions and open boundary conditions: as a matter of fact, in the former case the presence of $K_\mu^{(r)}(x)$ has no effect, while in the latter case the effect of $K_\mu^{(r)}(x)$ is to turn off the coupling of that link, which thus is not updated during the Monte Carlo (MC) evolution. Any intermediate choice $0 < c(r) < 1$ instead interpolates between these two extrema. The defect was chosen in all cases to be a $L_d\times L_d \times L_d$ cubic region set orthogonally to the $x_0=\tL-1$ time slice, and its position is effectively moved by translating the links of the periodic $c=1$ configuration in a random direction by one lattice spacing after each updating step.

\begin{figure}[!t]
\centering
\includegraphics[scale=0.4]{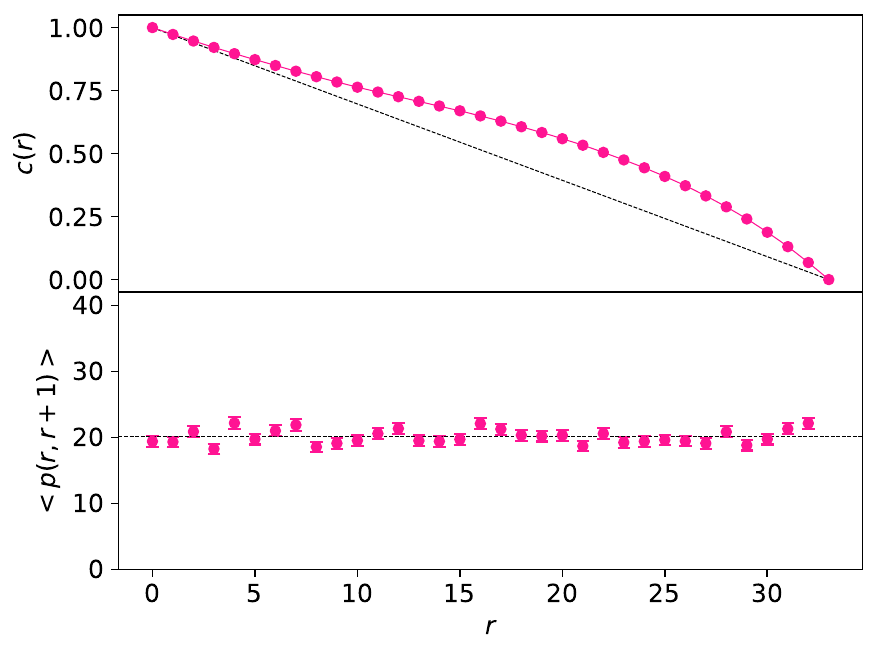}
\caption{Top panel: values of $c(r)$ adopted for the run with PTBC with $\tL=36$ using $N_r=34$ replicas and a defect size $L_d/a=6$, compared with a simple linear behavior in $r$. Bottom panel: corresponding mean swap acceptances $\braket{p(r,r+1)}\sim 20\%$.}
\label{fig:cr_ex}
\end{figure}

Each replica is updated independently using a $n_{\ov}$:1 mixture of over-relaxation and heat-bath local updating algorithms (this combination will be referred to in the following as ``standard algorithm'' or ``standard MC updating step''), and from time to time during the MC evolution a swap of configurations is proposed among adjacent replicas $(r,s)=(r,r+1)$, which is accepted via a standard Metropolis step with probability:
\beq
p(r,s) = \min\left\{1, \exp\left[-S_{\mathrm{W}}^{(r)}[U_s] -S_{\mathrm{W}}^{(s)}[U_r] +S_{\mathrm{W}}^{(r)}[U_r] +S_{\mathrm{W}}^{(s)}[U_s]\right] \right\},
\eeq
with $U_s$ and $U_r$ denoting the gauge configurations of the replicas $r$ and $s$ before the swap. The main idea of this algorithm is that the fast decorrelation of the topological charge obtained in the open replica, achieved because in this case $Q$ is no more constrained to assume integer values even in the continuum theory, is transferred towards the periodic replica thanks to the swaps. Moreover, one can now perform the measurement of the quantities of interest directly on the periodic replica (i.e., the one with $c=1$), where no unphysical contribution from the open boundary has to be taken into account. In order to improve the performances of the algorithm, the coefficients $c(r)$ have been tuned via short test runs in order to achieve roughly a constant swap acceptance $\braket{p(r,r+1)}\sim 20\%$. An example of the adopted values of $c(r)$ and of the related swap probabilities is shown in Fig.~\ref{fig:cr_ex}.

\FloatBarrier

\section{Results}\label{sec:res}

Simulations with and without PTBC have been performed on the set of lattices presented in Tab.~\ref{tab:simulation_points}. The choice of bare couplings and lattice sizes was made to obtain approximately the same physical size $\tl\equiv a \tL \sim 1.1$ fm in all cases. The results obtained in Ref.~\cite{Bribian:2021cmg} indicate that, with these parameters, the topological charge distribution of the $\tL=24$ lattice is correctly sampled by the standard algorithm, while the distribution observed for the large $\tL=36$ lattice, with a factor of 1.5 smaller lattice spacing, is almost frozen in the zero topological charge sector. Both points have also been simulated with the PTBC algorithm, using a number of replicas given by $N_r=18$ and 34 for $\tL=24$ and 36 respectively. As mentioned above, the size of the defect is kept fixed in physical units and has been set to 4 and 6 lattice spacings respectively. 

\begin{figure}[!t]
\centering
\includegraphics[scale=0.345]{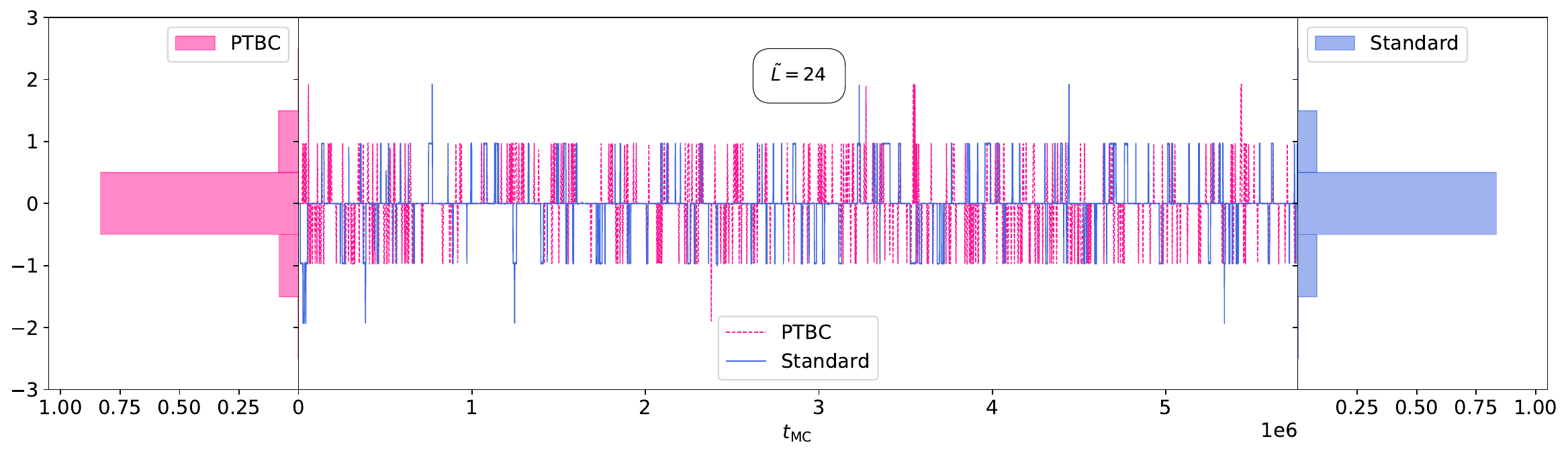}
\includegraphics[scale=0.345]{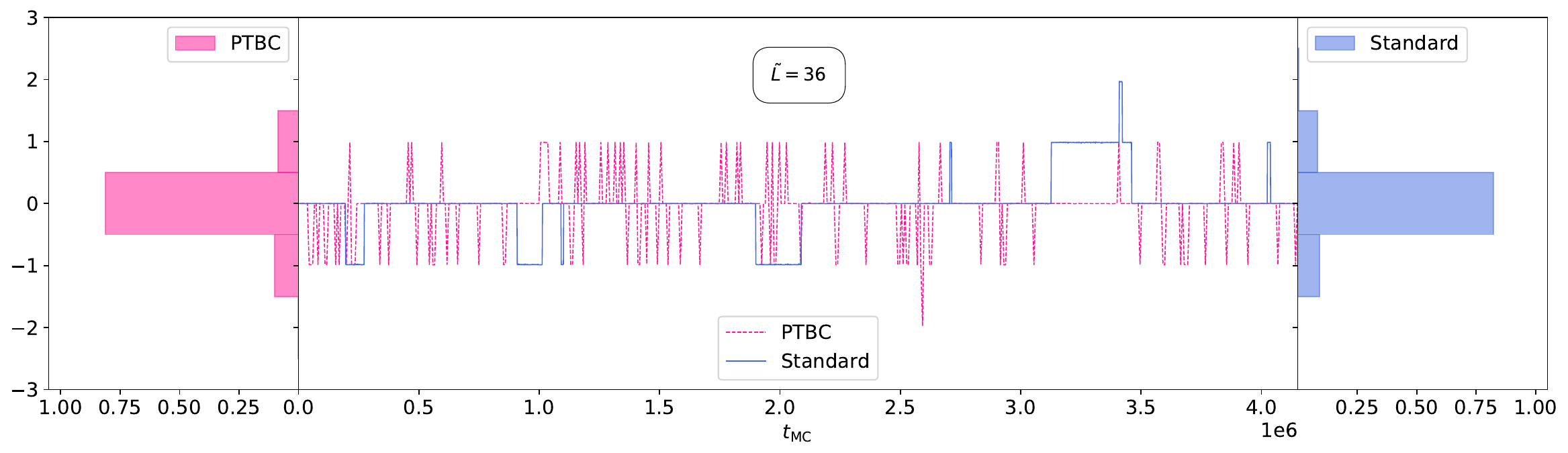}
\caption{Comparison of the MC histories of the topological charge and of the histograms of the topological charge distribution obtained using the PTBC (dashed lines) and the standard algorithm (solid lines). The time windows of the histories refers to a fraction of the collected sample for PTBC and to the whole collected sample for the standard algorithm, while the histograms refer to the full statistics in both cases. The Monte Carlo time is in both cases expressed in units of updating sweeps, meaning that the Monte Carlo time of both histories was multiplied by the number of over-relaxation sweeps $n_\ov$ and, in the case of PTBC, also by the number of replicas $N_r$. Top plot refers to $\tL=24$, bottom plot refers to $\tL=36$.}
\label{fig:story_Q}
\end{figure}

The Monte Carlo histories of the topological charge obtained with and without PTBC are shown in Fig.~\ref{fig:story_Q}, together with the histograms of the topological charge distribution. In both cases we expressed the Monte Carlo times in units of updating sweeps, meaning that we multiplied the Monte Carlo time of both algorithms by $n_{\ov}$ in order to keep into account the different number of over-relaxation sweeps employed in the two cases. Moreover, in the case of PTBC we also multiplied the Monte Carlo time by $N_r$ in order to keep into account the extra numerical effort required to update the replicas. The topological charge has been calculated using the clover discretization evaluated on flowed configurations, with the same value of the flow time used to determine the coupling, i.e. $t= c^2 \tl^2/8$ with $c=0.3$. Qualitatively, it can be observed that PTBC leads to an improved sampling of the topological charge in the case where the standard algorithm is affected by a significant topological freezing, cf.~Fig.~\ref{fig:story_Q} for the $\tL=36$ lattice\footnote{At this point a comment is in order: the lattice sizes used in step-scaling studies are relatively small in physical units and therefore dominated by the $Q=0$ sector, with the probability of exploring other charge sectors largely suppressed. This is a dynamical effect, not related to freezing, and is thus not addressed by PTBC. Other algorithmic strategies, such as multicanonical approaches~\cite{Berg:1992qua,Bonati:2017woi, Jahn:2018dke, Bonati:2018blm, Athenodorou:2022aay,Bonanno:2022dru}, can be used to improve the sampling of volume-suppressed topological sectors.}.

In the remainder of this section we perform a more quantitative analysis of the simulations with two main objectives in mind: first, to test whether the $Q=0$ sector is correctly sampled by the standard algorithm, and second, to determine the effect of parallel tempering by looking at various quantities, including estimates of auto-correlation times for the two types of algorithms.

\begin{table}[!t]
\begin{center}
\begin{tabular}{|c|c|c|c|c|}
\hline
$\tilde{L}$ & $\beta$ & $a/\sqrt{t_0}$ & $N_r$ & $L_d/a$ \\
\hline
24 & 6.4881 & 0.2770(35) & 18 & 4 \\
\hline
36 & 6.7790 & 0.1846(24) & 34 & 6 \\
\hline
\end{tabular}
\begin{tabular}{|c|c|c|c|c|c|c|}
\hline
$\tilde{L}$ & Algorithm & $N_r$ & $n_{\mathrm{meas}}$ & $n_{\ov}$ & $\Delta n_{\mathrm{meas}}$ & Total effort \\
\hline
\multirow{2}{*}{24} & PTBC & 18 & 9084 & 12 & 12 & 23.5M\\
\cline{2-7}
& Standard & & 10000 & 24 & 24 & 5.8M\\
\hline
\multirow{2}{*}{36} & PTBC & 34 & 3481 & 12 & 18 & 25.6M\\
\cline{2-7}
& Standard & & 3203 & 36 & 36 & 4.2M\\
\hline
\end{tabular}
\end{center}
\caption{Top table: Summary of simulation points, corresponding to simulating to 2 different values of the lattice spacing and the same lattice volume in physical units $\sim 1.1$ fm. The defect size was kept constant in physical units as well. The number of replicas $N_r$ was chosen in order to obtain a constant mean swap acceptance among adjacent replicas of about $20\%$. The lattice spacing is reported in terms of the well-known $t_0$ gradient flow scale. Bottom table: for both algorithms we report the number of measures $n_{\mathrm{meas}}$, the number of over-relaxation lattice sweeps per over-heat-bath lattice sweeps $n_\ov$, the separation between subsequent measures in terms of algorithmic steps $\Delta n_{\mathrm{meas}}$. In order to allow a fair comparison between the two algorithms, we also report the total numerical effort (in lattice sweeps) needed to generate each sample expressed in units of lattice sweeps, namely: $\text{total effort} = N_r \times n_{\mathrm{meas}} \times n_{\ov} \times \Delta n_{\mathrm{meas}}$ (where of course $N_r=1$ for the standard algorithm).}
\label{tab:simulation_points}
\end{table}

\begin{figure}[!t]
\centering
\includegraphics[scale=0.7]{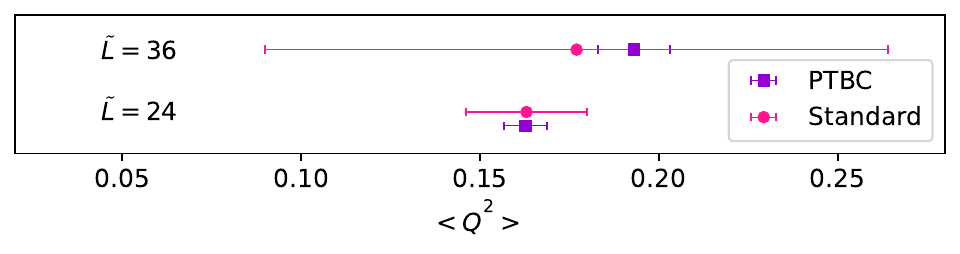}
\caption{Comparison of the obtained results for $\braket{Q^2}$ using the PTBC (square points) and the standard algorithm (round points). The topological charge was computed adopting the standard clover discretization on flowed configuration, for the same flow time employed to compute the coupling $t= c^2 \tl^2/8$ with $c=0.3$.}
\label{fig:comparison_q}
\end{figure}

A first quantitative manifestation of the effect of parallel tempering in the presence of topological freezing is shown in Fig.~\ref{fig:comparison_q}, where we compare the values of $\langle Q^2 \rangle$ obtained with and without PTBC (reported in Tab.~\ref{tab:res_summary}). With PTBC we obtain smaller relative errors on $\braket{Q^2}$ by a factor of 3 (10) for $\tL=24$ (36); however, we also have to keep into account that for PTBC we spent a larger numerical effort compared to the standard algorithm by a factor of 4 (6). This means that, for the $\tL=24$ lattice, where the standard algorithm is already capable of sampling correctly the topological charge distribution, the gain with PTBC is almost entirely compensated by the larger numerical effort spent, i.e., the two algorithms perform equally well. On the other hand, for the frozen simulation point corresponding to $\tL=36$, there is a clear gain in statistical accuracy using PTBC, which outperforms the standard algorithm. This is perfectly in line with our expectations, and confirms the qualitative behavior argued from the inspection of the MC histories of $Q$.

In Tab.~\ref{tab:res_summary} we also summarize our results for the TGF coupling, computed according to the formulas reported in~\cite{Bribian:2021cmg}, obtained with the two algorithms. In the case of $\lambda_{\TGF}$, results are given for the coupling averaged over all topological sectors and projected to the sectors with topological charge $Q=0$ and $Q=1$. The strong correlation between coupling and charge mentioned earlier can be easily seen from the large difference between the values of $\lambda_{\TGF}(Q=0)$ and $\lambda_{\TGF}(Q=1)$.

As it can be observed from the reported results, the same considerations done for $\braket{Q^2}$ apply also for the unprojected coupling $\lambda_{\TGF}(\text{No proj.})$. As a matter of fact, while for $\tL=24$ we see that the smaller error obtained with PTBC is exactly compensated by the larger numerical effort spent, for $\tL=36$ we find that PTBC yields a gain in terms of statistical accuracy of about a factor of 2 compared to the standard algorithm after keeping into account the different numerical efforts. This is perfectly in line with our expectations about the correlation between $\lambda$ and $Q$. Concerning instead the projected coupling, we observe no improvement in the relative errors adopting PTBC. Actually, we found that, in both cases, the measures of the projected coupling were essentially decorrelated. Thus, since for both algorithms we have the same number of measurements, errors were found in both cases to be of the same size.

The conclusions drawn so far are confirmed by our results for the auto-correlation times of the coupling ($\tau_\lambda$) and of the squared topological charge ($\tau_{Q^2}$)~\footnote{The auto-correlation times were computed from a standard binned jack-knife analysis.}, which are better-suited quantities to make a numerical-effort-independent comparison between the two algorithms.

More precisely, defining an ``effective'' auto-correlation time for the observable $\mathcal{O}$ as:
\beq\label{eq:tau_eff}
\tau^{(\mathrm{eff})}_{\mathcal{O}}= N_r \times \Delta n_{\mathrm{step}} \times n_{\ov} \times \tau_{\mathcal{O}},
\eeq
where of course $N_r=1$ for the standard algorithm, it is possible to make a fair comparison between the two adopted algorithms which automatically takes into account the different numerical efforts spent.

\begin{figure}[!t]
\centering
\includegraphics[scale=0.75]{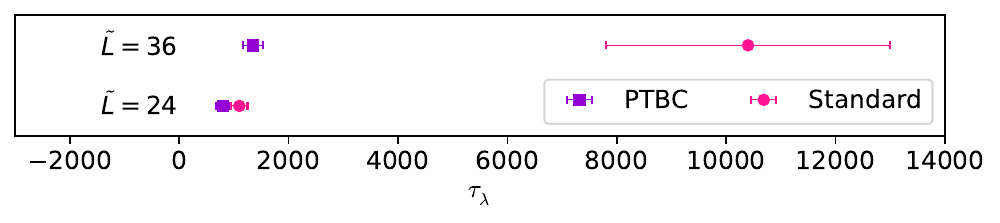}
\includegraphics[scale=0.75]{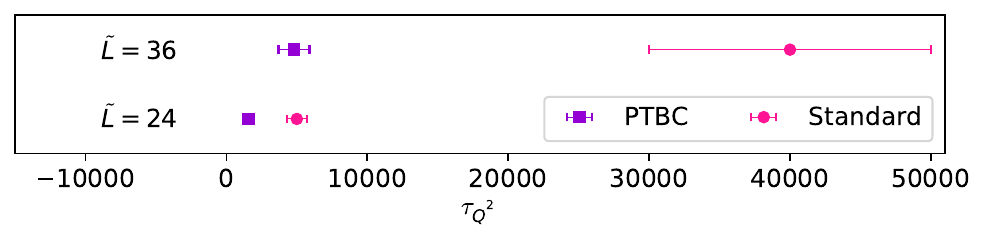}
\caption{Comparison of the obtained results for the auto-correlation times of $\braket{Q^2}$ (bottom panel) and of $\lambda_{\TGF}$ without projecting to a fixed topological sector (top panel), using the PTBC (square points) and the standard algorithm (round points). The auto-correlation time is in both cases expressed in units of lattice sweeps according to Eq.~\eqref{eq:tau_eff}.}
\label{fig:res_tau}
\end{figure}

Our results for the effective auto-correlation times are reported in Tab.~\ref{tab:res_summary_tau} and shown in Fig.~\ref{fig:res_tau}. The improvement achieved by using PTBC is clearly observed when looking at the auto-correlation time of $Q^2$ and of $\lambda_{\TGF}(\text{No Proj.})$ for $\tL=36$. Concerning the projected coupling, instead, in both cases we found our measures to be practically decorrelated, signalling that the auto-correlation time in these cases was smaller than the number of updating steps separating two subsequent measures. For this reason, in these cases we could just put upper bounds on $\tau$.

Given the results discussed so far, two main conclusions can be drawn:
\begin{itemize}
\item
A perfect agreement is found between the determinations obtained with and without PTBC when the coupling is projected onto the $Q=0$ sector. This points out that local topological fluctuations in the $Q=0$ sector are correctly sampled with the standard algorithm, even in the presence of a severe topological freezing.
\item
For the simulation point suffering from topology freezing, PTBC provides a very significant error reduction for both the square of the topological charge and the unprojected coupling, pointing out the expected strong correlation between $\lambda_\TGF$ and $Q$. For both simulation points we found very good agreement among the determinations of the unprojected coupling obtained with the two different algorithms.
\end{itemize}

\begin{table}[!t]
\begin{center}
\begin{tabular}{|c|c|c|c|c|c|}
\hline
$\tilde{L}$ & Algorithm & $\lambda_\TGF(\text{No Proj.})$ & $\lambda_{\TGF}(Q=0)$ & $\lambda_{\TGF}(Q=1)$ & $\braket{Q^2}$ \\
\hline
\multirow{2}{*}{24} & PTBC & 34.11(12) & 32.29(11) & 42.88(24) & 0.1627(60) \\
\cline{2-6}
& Standard & 34.01(20) & 32.18(10) & 42.92(25) & 0.163(17) \\
\hline
\multirow{2}{*}{36} & PTBC & 35.53(15) & 33.61(13) & 43.03(30) & 0.193(10) \\
\cline{2-6}
& Standard & 35.23(67) & 33.29(15) & 43.85(63) & 0.177(87) \\
\hline
\end{tabular}
\end{center}
\caption{Summary of the obtained results for $\braket{Q^2}$ and for the TGF coupling, with and without projecting to a fixed topological sector.}
\label{tab:res_summary}
\end{table}

\begin{table}[!t]
\begin{center}
\begin{tabular}{|c|c|c|c|c|c|}
\hline
$\tilde{L}$ & Algorithm & $\tau_\lambda(\text{No Proj.})$ & $\tau_\lambda(Q=0)$ & $\tau_\lambda(Q=1)$ & $\tau_{Q^2}$ \\
\hline
\multirow{2}{*}{24} & PTBC & 800(120) & $\lesssim$ 360(54) & $\lesssim$ 354(53) & 1570(230) \\
\cline{2-6}
& Standard & 1100(150) & $\lesssim$ 93(13) & $\lesssim$ 77(11) & 5000(700) \\
\hline
\multirow{2}{*}{36} & PTBC & 1350(180) & $\lesssim$ 940(130) & $\lesssim$ 910(240) & 4800(1100) \\
\cline{2-6}
& Standard & 10400(2600) & $\lesssim$ 200(30) & $\lesssim$ 940(230) & 40000(10000) \\
\hline
\end{tabular}
\end{center}
\caption{Summary of the obtained results for the auto-correlation times of $\braket{Q^2}$ and of the TGF coupling, with and without projecting to a fixed topological sector. The time is expressed in both cases expressed in units of lattice sweeps, see the definition in the text. The symbol $\lesssim$ denotes that the reported estimations of $\tau$ are just upper bounds, as the obtained measures of the corresponding observables were found to be decorrelated.}
\label{tab:res_summary_tau}
\end{table}

\begin{figure}[!t]
\centering
\includegraphics[scale=0.77]{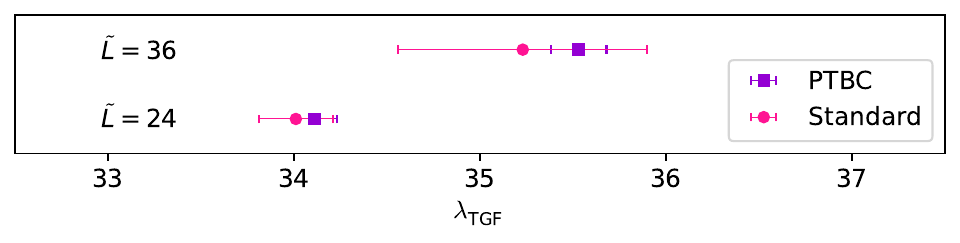}
\caption{Comparison of the obtained results for $\lambda_{\TGF}$ without projecting to a fixed topological sector using the PTBC (square points) and the standard algorithm (round points).}
\label{fig:res_obs}
\end{figure}

\FloatBarrier
\section{Conclusions}\label{sec:conclu}

In this proceedings we have presented a preliminary investigation of the TGF renormalized coupling of the $\SU(3)$ pure-gauge theory using the PTBC algorithm to improve the sampling of the lattice topological charge distribution.

The parallel tempering was shown to improve the auto-correlation time of the squared topological charge and of the unprojected coupling with respect to the previous
TGF calculation of Ref.~\cite{Bribian:2021cmg} for the simulation point affected by significant topological freezing.

On the other hand, in all cases the two algorithms seem to perform equally well concerning the projected coupling. Moreover, given that the PTBC algorithm always yields perfectly compatible results for the projected coupling when compared to the standard one, at this stage it appears that topological fluctuations within a specific topological sectors are well sampled, even in the presence of significant topology freezing.

The present preliminary investigation is part of a more extensive study, which will be the object of a forthcoming publication.

\section*{Acknowledgements}
This work is partially supported by the Spanish Research Agency (Agencia Estatal de Investigación) through the grant IFT Centro de Excelencia Severo Ochoa CEX2020-001007-S and, partially, by grant PID2021-127526NB-I00, both funded by MCIN/AEI/10.13039/501100011033. We also also acknowledge partial support from the project H2020-MSCAITN-2018-813942 (EuroPLEx) and the EU Horizon 2020 research and innovation programme, STRONG-2020 project, under grant agreement No 824093. Numerical calculations have been performed partially on the \texttt{Marconi} machine at Cineca, based on the agreement between INFN and Cineca, under projects INF22\_npqcd and INF23\_npqcd, and partially on the \texttt{Finisterrae~III} cluster at CESGA (Centro de Supercomputaci\'on de Galicia).

\providecommand{\href}[2]{#2}\begingroup\raggedright\endgroup

\end{document}